\documentclass[%
 reprint,
 amsmath,amssymb,
 aps,
]{revtex4-2}
\usepackage{amsmath}
\usepackage{tipa}
\usepackage{bbm}
\usepackage{txfonts}
\usepackage{graphicx}
\usepackage{dcolumn}
\usepackage{bm}
\usepackage{amssymb}
\usepackage{latexsym}
\usepackage{color}
\usepackage{ulem}
\usepackage[mathscr]{euscript}
\usepackage[colorlinks=true,linkcolor=blue,citecolor=blue,urlcolor=blue,]{hyperref}
\usepackage[doipre={doi:~}]{uri}

\begin{document}

\title{Magnetic monopole induced polarons in atomic superlattices}

\author{Xiang Gao$^1$} 
\author{Ya-Fen Cai$^1$} 
\author{Shao-Jun Li$^1$}
\author{Shou-Long Chen$^1$}
\author{Xue-Ting Fang$^1$}
\author{Qian-Ru Zhu$^1$}
\author{Lushuai Cao$^{1,2}$}\email[E-mail: ]{lushuai$\_$cao@hust.edu.cn}
\author{Peter Schmelcher$^{3,4}$}
\author{Zhong-Kun Hu$^{1,2}$}\email[E-mail: ]{zkhu@hust.edu.cn}
\affiliation{$^1$MOE Key Laboratory of Fundamental Physical Quantities Measurement $\& $ Hubei Key Laboratory of Gravitation and Quantum Physics, PGMF and School of Physics, Huazhong University of Science and Technology, Wuhan 430074, P. R. China \\
$^2$Wuhan Institute of Quantum Technology, Wuhan 430074, P. R. China\\
$^3$Zentrum für optische Quantentechnologien, Universität Hamburg, Luruper Chaussee 149, 22761 Hamburg, Germany \\
$^4$The Hamburg Centre for Ultrafast Imaging, Universität Hamburg, Luruper Chaussee 149, 22761 Hamburg, Germany
}


\begin{abstract}
Magnetic monopoles have been realized as emergent quasiparticles in both condensed matter 
and ultracold atomic platforms, with growing interests in the coupling effects between the monopole
and different magnetic quasiparticles. 
In this work, interaction effects between monopoles and magnons are investigated for an atomic pseudospin chain. 
We reveal that the monopole can excite a virtual magnon cloud in the paramagnetic chain, thereby
giving rise to a new type of polaron, the monopole-cored polaron (McP).
The McP is composed of the monopole as the impurity core and the virtual magnon excitation as the dressing cloud.
The magnon dressing facilitates the Dirac string excitation and impacts the monopole hopping. 
This induces an anti-trapping effect of the McP, 
which refers to the fact that the dressing enhances the mobility of the McP, 
in contrast to the self-trapping of the common polarons.
Moreover, heterogeneous bipolarons are shown to exist under the simultaneous doping of a north
and a south monopole. The heterogeneous bipolaron
possesses an inner degree of freedom composed of two identical impurities.
Our investigation sheds light on the understanding of how the coupling between the impurity core
and the dressing cloud can engineer the property of the polaron.
\end{abstract}

\maketitle

\section{Introduction}
\label{section:I}
The quantum magnetic monopole, firstly introduced by P. M. Dirac to explain the quantization of the electric charge
 \cite{dirac1931quantised,Milton_2006},
has been realized as an emergent quasiparticle on different platforms, 
such as condensed-matter (artificial) spin ices
\cite{Castelnovo2008Magnetic,Fennell2009Magnetic,Jaubert2009Signature,Morris2009Dirac,King2021Qubit} 
and ferromagnetic films \cite{Zhong2003Anomalous,Guang2021Superposition},
as well as ultracold atomic Bose-Einstein condensates 
\cite{Savage2003Dirac,Pietila2009Creation,Pietila2009NonAbelian,Tiurev2016Decay,Li2017Dirac,Tiurev2019Creation,Mithun2022Existence,Ruokokoski2011Ground,Ray2014Observation,Ray2015Observation,Ollikainen2017Experimental,sugawa2018second,Ollikainen2019Decay,Gao2022Interaction}
and pseudospin chain \cite{Gao2022Interaction}.
The two fingerprints of the monopole, namely the singular magnetic field and the Dirac string, have been identified
and stimulated emerging interests in investigating how these fingerprints can induce unique coupling between the monopole
and different magnetic quasiparticles. For instance,
it has been found that the monopole can be attracted to and form bound states with the intrinsic quasiparticle of ferromagnetic
spin lattices, i.e. the ferromagnetic domain wall \cite{Guang2021Superposition} and magnetic kink \cite{Gao2022Interaction} 
in the two- and one-dimension spin lattices, respectively. It is intriguing to turn to the monopole
coupling to other quasiparticles, e.g. the magnon, 
and investigate new effects and phenomena arising from the coupling to the monopole.

It has been well known that the impurities doped to a host environment can excite and get bound to the 
intrinsic excitations of the host, and form the emergent quasiparticle of polarons. The concept of polaron was first
introduced in condensed matter physics \cite{landau1933bewegung}, and soon found applications in diverging fields 
\cite{BISHOP1973391, Kutschera1993Proton, de1997evidence, Bredas1985Polarons, Esther2005Charge, Gershenson2006Colloquium}. Ultracold atomic gases provide a flexible platform for the investigation of various types of polarons,
such as the Bose \cite{Catani2012Quantum, Hohmann2015Neutral, Jorgensen2016Observation, Grusdt2016Condensate, Camacho2018Bipolarons, Dehkharghani2018Coalescence, Will2021Polaron, Hu2016Bose, Yan2020Bose, Christensen2015Quasiparticle, Shchadilova2016Quantum, Mistakidis2019Effective, Christos2019Two, Ardila2020Strong, Astrakharchik2021Ionic, Fabian2021Impurities, Mistakidis_2022Inducing, Richard2022Selfstabilized}, 
Fermi \cite{Schirotzek2009Observation,Kohstall2012Metastability,Koschorreck2012Attractive,Zhang2012Polaron,Cetina2015Decoherence,Cetina2016Ultrafast,Scazza2017Repulsive,Tylutki2017Coherent,Huber2019Inmedium,  Christos2019Two, Dean2021Impurities} 
and magnetic polarons \cite{Fukuhara2013Quantum,Grusdt2018Parton, Fabian2018Meson,Chiu2019String,Koepsell2019Imaging,Koepsell2021Microscopic}, 
which are mainly specified by the host environment and the corresponding virtual dressing cloud.
The Bose polarons normally refer to the polarons generated in the host environment of Bose gases, 
where the impurities, either bosons 
\cite{Catani2012Quantum, Hohmann2015Neutral, Jorgensen2016Observation, Grusdt2016Condensate, Camacho2018Bipolarons, Dehkharghani2018Coalescence, Will2021Polaron} 
or fermions \cite{Camacho2018Bipolarons,Dehkharghani2018Coalescence,Hu2016Bose,Yan2020Bose,Will2021Polaron}, 
get bound to the virtual phonon excitations. In the Fermi polaron, the host environment changes to the Fermi gas, 
and the dressing cloud becomes the particle-hole excitations. Besides the Bose and Fermi polarons,
there exist a big family of magnetic polarons, of which the magnetic impurities \cite{Fukuhara2013Quantum}  or charged \cite{Grusdt2018Parton, Fabian2018Meson, Chiu2019String, Koepsell2019Imaging, Koepsell2021Microscopic} are doped to spin lattices residing in different magnetic phases, 
such as the (anti-)ferromagnetic and paramagnetic phases.
The magnetic polarons hold the key to various important open questions in condensed matter physics, such
as the high-$T_c$ superconductivity and spin transports \cite{mott1990spin}, 
and have become a key interest in ultracold atomic physics. 

In this work, we bring in a new member to the magnetic polaron family, dubbed the monopole-cored polaron (McP),
which is formed by the magnetic monopole as the impurity core and the virtual magnon excitation as the dressing cloud.
The virtual magnon is excited by the singular magnetic field of the monopole. 
The interplay between the magnon dressing cloud and the Dirac string of the monopole, encapsulated in the spin flipping along the hopping path of the monopole, further induces the anti-trapping effect to the McP. 
The anti-trapping effect means that the dressing magnon cloud enhances the mobility of the 
McP, and a stronger excitation of the magnon cloud leads to a higher mobility, which is in contrast to the well-known
self-trapping effect of the existing polarons \cite{Schirotzek2009Observation,Koschorreck2012Attractive,Christensen2015Quasiparticle,
Grusdt2016Condensate,Scazza2017Repulsive,Ardila2020Strong}.
When simultaneously doping one north and one south monopole to the paramagnetic (pseudo-)spin chain,
the heterogeneous bipolaron with the two monopoles as the impurity cores are generated, 
of which the heterogeneity is attributed to the distinguishability of the two cores.
The heterogeneous bipolaron possesses an extra inner degree of freedom, which could find potential application
in manipulating the bipolaronic transport processes \cite{Bredas1985Polarons, mott1990spin, Alexandrov1994Bipolarons}.

This work is organized as follows: 
In section \ref{section:II}, we introduce the setup and pseudospin mapping.
In section \ref{section:III}, we demonstrate the generation and dynamics of single McP.
Section \ref{section:IV} contains an investigation of the coupling effects between two McPs induced by south and north magnetic monopoles. 
A brief discussion and outlook are given in Section \ref{section:V}.

\section{Set up and Hamiltonian}
\label{section:II}
The atomic pseudospin chain considered in this work is based on a one-dimensional (1D) atomic dipolar superlattice gas (DSG),
which consists of spin-polarized fermions confined in the 1D double-well superlattice with periodic boundary condition. 
The fermions interact with each other by repulsive dipole-dipole interaction (DDI) \cite{Lahaye_2009Thephysics}.
Under the tight-binding approximation, the DSG system can be described by the following Fermi-Hubbard Hamiltonian:
\begin{equation}
    \begin{aligned}
    {\hat H_{\rm{FH}}}=  &- J\sum\limits_{i}^M {\left( {\hat f_{2i}^\dag {\hat f_{2i - 1}} + {\rm{H}}.{\rm{c}}.} \right)} 
     - {J_1}\sum\limits_{i}^{M - 1} {\left( {\hat f_{2i}^\dag {{\hat f}_{2i + 1}} + {\rm{H}}.{\rm{c}}.} \right)}  \\
      &+ \sum\limits_{i < j}^{2M} {{V_d}\left( {i,j} \right){{\hat n}_i}{{\hat n}_j}},\label{Eq:1}
    \end{aligned}
    \end{equation}
    \label{eq1}
\noindent
where $\hat f_{2i-1/2i}^{(\dag)}$ annihilates (creates) a fermion in the left/right site of the $i$-th supercell,
with ${\hat n_i} \equiv \hat f_i^\dag {\hat f_i}$ denoting the occupation-number operator. 
The first two terms in ${\hat H_{\rm{FH}}}$  describe the intra- and inter-cell hopping, respectively,
with $J\gg J_1$. The last term refers to the dipole-dipole interaction between two fermions located in sites $i$ and $j$,
with the interaction strength approximated by ${V_d}\left( {i,j} \right) =  D/\left| x_i-x_j \right|^3$, where 
$x_{i(j)}$ refers to the local minimum of the $i(j)$-th site, and D is the dipole-dipole interaction strength.
In our investigation, we take $J = 50{J_1}$ and truncate the $V_{d}(i,j)$ to the nearest neighbor (NN) interaction, 
where the intra- and inter-supercell NN interaction takes the value of $V_d \left( {2i-1,2i} \right)=1.5d$ and 
$V_d \left( {2i,2i+1} \right)=d$. The main results of this work are, however, not dependent on the quantitative
settings, and can be generalized to a broad parameter region.


\begin{figure}[t]
	\centering
	\includegraphics[trim=0 20 0 -20, width=0.48 \textwidth]{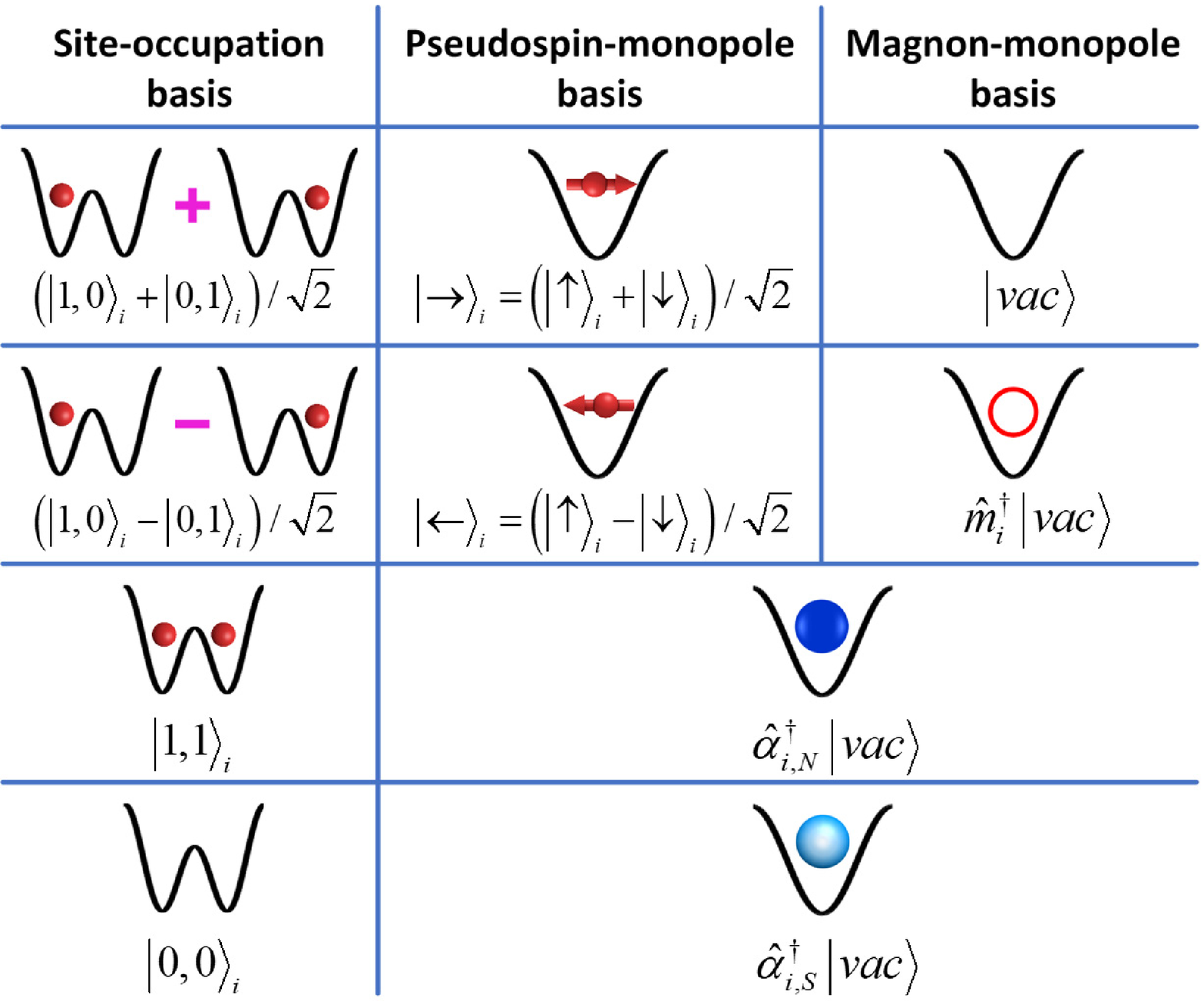}
	\caption{
    \label{fig1}
    The one-to-one correspondence between occupation states (left column) and pseudospin-monopole states (middle column),
    as well as the one that between pseudospin-monopole states and magnon-monopole states (right column).
    The red hollow circle refers to the magnon, and the dark and light blue balls sketch the south magnetic monopole (NM) and north magnetic monopole (SM), respectively.
	}
\end{figure}

The DSG system can be transformed into a pseudospin chain through the mapping from the local
occupation to the pseudospin states,
which has been widely applied to atoms confined in a tilt potential \cite{Sachdev2002Mott, simon2011quantum, Meinert2013Quantum, Buyskikh2019Spin}, double-well 
\cite{Li2016SpinOrbit,li2017stripe,Yin2015Magnetic,CAO2017303Collective,Gao2022Interaction} 
and ladder lattices \cite{Yongyao2017Long}. 
In the standard pseudospin mapping for the double-well superlattice, the single-occupation states
of each cell, e.g. the $i$-th cell, are mapped to the pseudospin states, as 
$\left| 1,0 \right\rangle_i\rightarrow\left|  \uparrow  \right\rangle_i$
and $\left| 0,1 \right\rangle_i\rightarrow\left|  \downarrow  \right\rangle_i$,
where  $\left| n_L, n_R \right\rangle_i$ denotes the occupation states in the DSG chain 
with $n_L$ and $n_R$ atoms occupying the left and right sites of the $i$-th cell, respectively,
and $|\uparrow (\downarrow)\rangle_i$ refers to the states of the $i$-th site of the pseudospin chain. 
In \cite{Gao2022Interaction}, we have further demonstrated that the double- and null-occupations 
$\left| 1,1 \right\rangle_i$ and $\left| 0,0 \right\rangle_i$ can be mapped to the north and south
magnetic monopoles, respectively. In this work, we adapt the pseudospin mapping to the 
magnon-monopole basis on top of the paramagnetic phase, with all pseudospins aligning to the 
same direction $\prod_{i=1}^M\left| \rightarrow  \right\rangle_i$. The intrinsic excitation
of the paramagnetic phase is the magnon, and a magnon refers to one spin flipping
to the opposite direction $|\leftarrow\rangle$. 
In the following, we take the paramagnetic state $\prod_{i=1}^M\left| \rightarrow  \right\rangle_i$
as the vacuum state, on top of which $\left| \leftarrow  \right\rangle_i$,
$\left| 1,1 \right\rangle_i$ and $\left| 0,0 \right\rangle_i$ are taken as the excitation of a magnon,
a north and south monopole on the $i$-th site, respectively.
Figure 1 sketches the pseudospin mapping of the DSG chain, in which
the one-to-one mapping between the site-occupation states of the DSG chain to the pseudospin-monopole 
and magnon-monopole basis is illustrated.

Following the pseudospin mapping, ${\hat H_{\rm{FH}}}$ can be transformed to the magnon-monopole basis, as:
\begin{widetext}
\begin{subequations}
        \begin{align}
            {\hat H_{{\rm{eff}}}} = &{\hat H_{{\rm{mgn}}}} + {\hat H_{{\rm{mpl}}}} + {\hat H_{{\rm{mgn-mnp}}}},  \\
            {\hat H_{\rm{mgn}}} =  &2J\sum\limits_{i}^M {\hat n_{i,M}}  - \frac{d}{4}\sum\limits_{i}^{M - 1} {\left( {\hat m_i ^\dag {{\hat m}_{i  + 1}} + {\rm{H.c.}}} \right)} 
            - \frac{d}{4}\sum\limits_{i}^{M - 1} {\left( {\hat m_i ^\dag \hat m_{i  + 1}^\dag  + {\rm{H.c.}}} \right)},   \\
            {\hat H_{\rm{mnp}}} =& - \frac{d}{2}\sum\limits_i^M {{{\tilde \sigma }_{z,i}}}  +2d\sum_i^{M} {\hat n_{i,S}}
            + \frac{d}{4}\sum\limits_i^{M-1} {{{\tilde \sigma }_{z,i}}{{\tilde \sigma }_{z,i + 1}}}
            - \frac{J_1}{2}\sum\limits_{i,\Lambda}{\left( \hat \alpha_{\Lambda,i}^\dag \hat \alpha_{\Lambda,i+1} +\rm{H.c.} \right)}
            -\frac{J_1}{2}\sum\limits_{i,\Lambda_1 \neq \Lambda_2}^{M-1}{ \left( \hat \alpha_{\Lambda_1,i} \hat \alpha_{\Lambda_2,i+1}+\rm{H.c.}\right)},  \\ 
            {\hat H_{{\rm{mgn-mnp}}}} = &- \frac{d}{4}\sum\limits_i^M \left\{ { {{{\tilde \sigma }_{z,i}}} \left[ {\left( {\hat m_{i  - 1}^\dag  + {{\hat m}_{i  - 1}}} \right) - 							\left( {\hat m_{i  + 1}^\dag  + {{\hat m}_{i  + 1}}} \right)} \right]} \right\} 
            - \frac{{{J_1}}}{2}\sum\limits_{i,\Lambda }^{M - 1} \left\{ {\hat \alpha _{\Lambda ,i}^\dag {{\hat \alpha }_{\Lambda ,i + 1}}\left[ {\left(-1 \right)^{\frac{\tilde \sigma_{z,i + 1} -1}{2} }\left( {{{\hat m}_i} - \hat m_{i + 1}^\dag } \right)}+ {{\hat m}_i}\hat m_{i + 1}^\dag  \right] + {\rm{H.c.}}}  \right\}   \\ 
            &- \frac{{{J_1}}}{2}\sum\limits_{i,\Lambda_1  \ne \Lambda_2}^{M - 1} \left\{ {{{\hat \alpha }_{\Lambda_1 ,i}^{\dag}}{{\hat \alpha }_{\Lambda_2 ,i + 1}^{\dagger}}\left[ {\left(-1\right)^ {\frac{\tilde \sigma_{z,i+1}+1}{2}} \left( {\hat m_i  + \hat m_{i + 1} } \right)} + \hat m_i \hat m_{i + 1} \right] + {\rm{H.c.}}} \right\}. \notag
        \end{align}
        \label{eq2}
\end{subequations}
\end{widetext}
\noindent
In $\hat H_{{\rm{eff}}}$, $\hat m_i ^{(\dag)}$ and $ {\hat \alpha_{i,\Lambda}^{(\dag)}}$ denote the creation
(annihilation) of a magnon and $\Lambda$-type monopole in the $i$-th site, respectively,
where $\Lambda=S(N)$ denotes the south (north) monopole.
The magnon number operator is defined as $\hat n_{i,M} = \hat m_i ^\dag {\hat m_i }$, and the Pauli-like 
operator for monopoles is also introduced as
${{{\tilde \sigma }_{z,i}}}=\hat n_{i,S}-\hat n_{i,N}$, 
where $\hat n_{i,\Lambda} = \hat \alpha_{i,\Lambda}^\dag {\hat \alpha_{i,\Lambda}}$ is the number operator of the monopole.

${\hat H_{\rm{mgn}}}$ describes the transverse Ising spin chain in the magnon picture.
The transverse magnetic field, i.e. the first term of ${\hat H_{\rm{Mag}}}$, originates from intra-cell
hopping of the DSG system and determines the chemical potential of the magnon. The dipole-dipole interaction of
the DSG system gives rise to the Ising type spin-spin interaction, as well as the hopping and pair 
creation/annihilation of the magnon, which is related to the last three terms of ${\hat H_{\rm{mgn}}}$, respectively.
${\hat H_{{\rm{mnp}}}}$ further describes the monopole behavior in the atomic pseudospin chain.
The first plus second and the third terms of ${\hat H_{{\rm{mnp}}}}$ refer to the chemical potential and the nearest-neighbor interaction
of the monopoles, which both originates from the dipole-dipole interaction of
the DSG chain. The inter-cell hopping leads to the hopping and pair creation/annihilation of the monopoles, 
given by the last two terms of ${\hat H_{{\rm{mnp}}}}$.

The coupling between the monopole and magnon, i.e. the focus of this work, is described by ${\hat H_{{\rm{mgn-mnp}}}}$,
which is induced by the singular magnetic field and the Dirac string of the monopole.
The singular magnetic field around the monopole can polarize neighbor spins and induces the creation and annihilation of the magnon, which is given by the first term of ${\hat H_{{\rm{mgn-mnpl}}}}$. 
The hopping of the monopole can excite the Dirac string, which is manifested as the spin flipping along the 
hopping path of the monopole 
\cite{Castelnovo2008Magnetic,Fennell2009Magnetic,Jaubert2009Signature,Morris2009Dirac,Gao2022Interaction}.
In the magnon-monopole picture, the Dirac string excitation becomes the excitation/annihilation and 
counter-propagation of the magnon along with the monopole hopping, 
which is described as the second term of ${\hat H_{{\rm{mgn-mnp}}}}$.
The last term ${\hat H_{{\rm{mgn-mnp}}}}$ indicates that a pair of a north and a south monopole 
can be excited by the annihilation of a single and a pair of magnons. We will reveal
the rich coupling phenomena induced by $\hat H_{{\rm{mgn-mnp}}}$ in the following.

\begin{figure}[t]
	\centering
	\includegraphics[trim=10 20 0 -5,width=0.48 \textwidth]{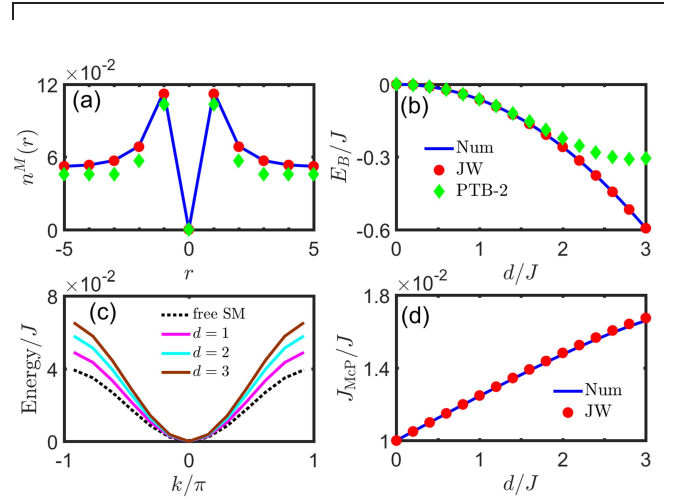}
	\caption{
	\label{fig2}
	(a) The magnon density around the SM obtained from the numerical calculation (blue solid lines) and the McP basis ansatz
        with Jordan-Wigner transform (red solid circles) and second-order perturbation treatment (green solid diamonds) for $d=2.5$ in the 11-sites pseudospin chain.
    (b) The binding energy $E_B$ as a function of the monopole (spin)-spin interaction strength $d$.
	(c) The dispersions of a free SM (black dashed line) and the McP induced by the SM for $d=1$ (purple solid line), $2$ (cyan solid line) and $3$ (brown solid line).
	(d) The hopping amplitude of McP $J_{\rm{McP}}$ as the function of $d$.
    }
\end{figure}

\section{The magnetic polarons: generation, dynamics and interaction}
\label{section:III}
In this section, we focus on the doping of a single monopole to the periodic paramagnetic 
pseudospin chain and demonstrate the generation and dynamical properties of the McP. 
The generation of the McP through the singular magnetic field of the monopole is derived
by the Jordan-Wigner transformation \cite{Lieb1961Two,Hu2021First}, 
which takes into account multiple magnon
excitations and provides an accurate description of the formation and dynamical evolution of
the McP in a wide range of interaction strength.

\subsection{The formation of the McP}
The McP is composed of the monopole as the impurity core and the dressing virtual magnon cloud,
which is excited by the singular magnetic field around the monopole. The McP basis state
$\left| i \right\rangle _{\Lambda-\rm{McP}}$, denoting the McP with a $\Lambda$ monopole core centered at the $i$-th site, 
can be obtained by the Born-Oppenheimer approximation \cite{Will2021Polaron}, 
in which $\left| i \right\rangle _{\Lambda-\rm{McP}}$ is taken as the ground state of 
the pseudospin chain with the monopole localized at the $i$-th site, as 
\begin{equation}
    \begin{aligned}
        \left| i \right\rangle _{\Lambda-\rm{McP}} = \left| i \right\rangle_{\Lambda} \otimes \left| \psi \right\rangle_{i, \Lambda},
    \end{aligned}
    \label{eq3}
\end{equation}
where $\left| i \right\rangle _{\Lambda} = \hat \alpha_{\Lambda,i}^\dagger \left| vac \right\rangle$, and
${\left| \psi \right\rangle _{i, \Lambda}} $ is the ground state of the remaining pseudospin chain.
The derivation of ${\left| \psi \right\rangle _{i, \Lambda}} $ based on the Jordan-Wigner transformation \cite{Lieb1961Two,Hu2021First}
is provided in Appendix.\ref{App1}, and we also compare the results to that obtained by second-order perturbation (Appendix.\ref{App3}).

The effective Hamiltonian spanned in the basis $\{\left| i \right\rangle^\Lambda_{\rm{McP}}\}$ reads:
\begin{equation}
    \begin{aligned}
{\hat H_{\rm{McP}}} = &-\sum\limits_{i}^{M - 1} {{J_{\rm{McP}}}\left( {\hat p_{i}^\dag {{\hat p}_{i+1}} + \rm{H.c.}} \right)
+E_B},
    \end{aligned}\label{Eq4}
\end{equation}
where ${\hat p_{i}^\dag}$ and ${\hat p_{i}}$ are the creation and annihilation operators of McP, respectively.
The first term in ${\hat H_{\rm{McP}}}$ describes the hopping of the McP, with the hopping strength $J_{MP}$
dependent on the coupling strength between the monopole and magnon.
The constant energy offset $E_B$ refers to the binding energy of the McP, and measures the 
energy difference between the McP and a bare monopole with no magnon cloud dressing.

Taking the doping of a south monopole for instance, we present in Figs. 2(a) and (b) the density 
distribution of the magnon and the binding energy of the McP, respectively. 
Figure 2(a) shows the magnon density
$ n^M(r)=\sum_x \left\langle\hat n_{x,S} \hat n_{x+r,M}\right\rangle$ as a function of the relative
distance $r$ between the magnon and the monopole, with $<.>$ indicating the expectation with respect to the ground state.
$ n^M(r)$ in Fig. 1(a) visualizes the virtual excitation of the magnon cloud around the monopole, 
which is peaked at the nearest neighbor site of the monopole and decays as being away from the monopole.
The total density of the magnon excitation is far below unity, 
which reflects the virtual-excitation nature of the dressing magnon cloud.
Figure 2(b) plots the binding energy $E_B$ as a function of the dipolar interaction 
strength $d$, which determines the coupling between the monopole and the dressing magnon cloud. 
We observe, for one thing, that the binding energy is always negative, 
indicating that the formation of the McP lowers the total energy and is energetically more favorable
than the bare monopole in the paramagnetic (pseudo)spin chain. 
For another, $E_B$ decreases as $d$ increases, and it costs more energy to decouple the monopole and dressing
magnon cloud in the McP state. In Figs. 2(a) and (b) we particularly 
compare the numerical results obtained by the original Hamiltonian of $\hat H_{\rm{FH}}$,
to the analytical results by the Jordan-Wigner transformation and the second-order perturbation approaches.
It can be found that the Jordan-Wigner prediction matches the numerical results well,
but the perturbation theoretical prediction deviates from the numerical results in the strong interaction regime.
Given that the Jordan-Wigner transformation takes into account all possible magnon excitation channels, 
while the second-order perturbation truncates to four-magnon excitations,
the comparison reveals that the dressing cloud of the McP contains more than four-magnon excitations 
in the strong interaction regime, even though the total magnon density is weak. 
This is consistent with the Bose polaron picture,  in which the multi-phonon excitation also plays an important role in the
polaron formation \cite{Shchadilova2016Quantum,Will2021Polaron}.

\begin{figure}[t]
\centering
	\includegraphics[trim=15 15 5 -10,width=0.48 \textwidth]{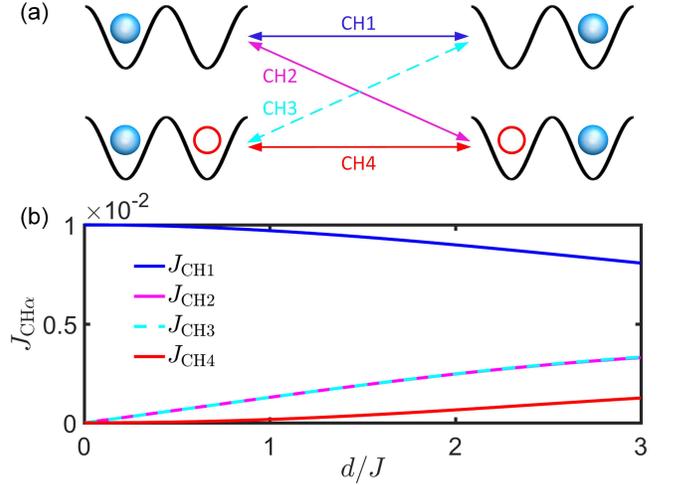}
	\caption{
    (a) The four possible hopping channels (CH1-CH4) of the SM: 
    CH1 refers to the tunneling of the SM on the magnon vacuum background.
    CH2 and CH3 represent the tunneling of the SM with magnon creation and annihilation, 
    and the counter-propagating channel between the SM and the magnon is illustrated in CH4.
    (b) Contributions of the CH1 (blue solid line), CH2 (purple solid line), CH3 (cyan dashed line), and CH4 (red solid line) to the hopping strength of the McP.
	\label{fig3}
    }
\end{figure}

Figure 2(c) further presents the dispersion relation of the McP, with a comparison to that
of the bare monopole in the DSG chain. The bandwidth of the McP, as shown in Fig. 2(c) is wider
than that of the bare monopole, and increases with the interaction strength $d$. This behavior indicates that
the mobility of the McP is higher than that of the bare monopole, and gets even stronger as the 
coupling between the monopole and magnon increases. This is in contrast to the well-known self-trapping
effect of the 'traditional' polarons, of which the mobility decreases as the coupling between the impurity
core and the dressing cloud increases
\cite{Schirotzek2009Observation, Koschorreck2012Attractive, Christensen2015Quasiparticle,
Grusdt2016Condensate,Scazza2017Repulsive,Ardila2020Strong}.
We term this phenomenon as the anti-trapping effect, which manifests itself 
as a major difference between the McP and the magnetic polaron for 'normal' magnetic impurities.
In order to more explicitly confirm the anti-trapping effect, we show the hopping strength of the McP 
as a function of $d$, as shown in Fig. 2(d), which indeed presents a monotonous increase with $d$.

The mechanism of the anti-trapping effect can be attributed to the Dirac string of the monopole, 
which refers to the spin flipping along the monopole hopping path. The excitation of the 
Dirac string leads to multiple hopping channels of the monopole, as sketched in Fig. 3(a),
which include the direct hopping of the monopole on top of the paramagnetic vacuum marked as CH1, 
as well as the hopping of the monopole associated with the excitation, annihilation, and counter-propagating 
magnon in the nearest neighbor site, marked as CH2, CH3, and CH4, respectively, in Fig. 3(a).
The channel of CH1 corresponds to the third term of $\hat H_{\rm{mnp}}$, and channels of CH2,3,4 are included in
the second term of $\hat H_{\rm{mgn-mnp}}$. These hopping channels contribute differently to the hopping 
strength of the monopole, as shown in Fig. 3(b), where the hopping strength through CH1 decreases with $d$, 
while the hopping in the remaining channels increases with $d$, leading to the anti-trapping effect. 
Figure 3(b) confirms that the anti-trapping effect is attributed to the Dirac string excitation of
the monopole along hopping.
The intuitive picture of the anti-trapping effect can then be understood as follows: 
In the paramagnetic phase, the hopping of the McP, as in the case of 'normal' polarons, disturbs the dressing magnon distribution 
around the monopole  with an energy cost, and effectively creates an energy barrier for the hopping.
The Dirac string excited along the monopole hopping, however, can simultaneously rearrange the magnon distribution around the monopole, and suppresses the energy barrier. 
As a comparison, the normal polarons can only hop along the chain through the channel of CH1 
and experience the energy barriers induced by the dressing cloud, which results in the 
self-trapping effect.
The anti-trapping effect of McP also illustrates how the coupling between the impurity core and the dressing cloud can modify the dynamical property of the corresponding polaron.

\subsection{Dynamics of the McP} 
It is necessary to test whether the McP picture can well capture the  
dynamical evolution of the paramagnetic (pseudo)spin chain with the monopole doping.
For this purpose, we further investigate the dynamical evolution of the McP subjected to a gradient magnetic field.
The gradient field exerted to the
atomic pseudospin can be introduced by the tilt potential added to the double-well superlattice, 
and we focus on the quadratic tilt 
of the form $\hat V^{\rm{grad}}_{\rm{FH}}=\sum_i(g_1\times i+g_2\times i^2) \hat{f}^{\dagger}_i\hat{f}_i$, of which
$g_1$ and $g_2$ denote the strength of the linear and quadratic tilt, respectively.
The tilt potential introduces an effective magnetic field on the pseudospin, and in the McP picture
it turns out to be:
\begin{subequations}
    \begin{align}
    &\hat V_{\rm{S-McP}}=\sum_j^M {\left[ {- 2\left( {{g_1} - {g_2}} \right) \times j - 4{g_2} \times {j^2}} + \mathcal{O}(j) \right]} \hat p_j^\dagger \hat p_j ,\\
    &\mathcal{O}(j)= _{\rm{McP}} {\left\langle j \left| \sum_{i \neq j} ( - 2{g_2} \times i)(\hat m_i^\dagger + \hat m_i) \right| j \right\rangle} {}_{\rm{McP}} \sim g_2^2.
    \end{align}
\end{subequations}
In Eq. (5a), $\hat V_{\rm{S-McP}}$ is dominated by the response of the monopole to the magnetic field, 
which is exerted by both the linear and quadratic gradient potential to the McP. 
$\mathcal{O}(j)$ denotes the minor correction from the effect of the magnetic field to the magnon cloud, 
since the total density of the magnon cloud is relatively weak.

\begin{figure}[t]
\centering
	\includegraphics[trim=10 20 -20 0,width=0.5 \textwidth]{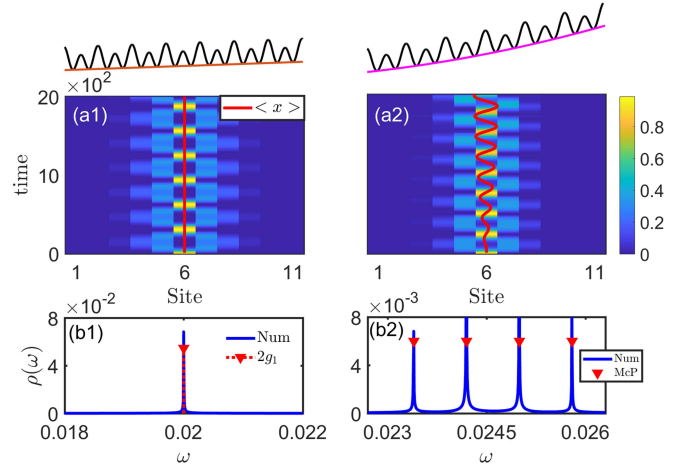}
	\caption{
	\label{fig4}
	Temporal evolution of the one-body density of McP for linear gradient fields with $(g_1,g_2)=(J/100,0)$ (a1) and quadratic gradient fields ($g_1=J/100$, $g_2=g_1/100$). 
    The red solid lines show $\left\langle x\right\rangle$ as the function of time.
    The top panel of (a1) and (a2) shows the superlattice potential under a first-order and second-order gradient field.
    (b1) and (b2) The density spectrum for evolution time $T= 10^6 \frac{1}{J}$.
    In (b1), the standard Bloch oscillation (BO) occurs of McP with $\omega=2g_1$ under linear gradient fields.
    (b2) shows the quadratic BO feature of McP, with the center frequency of $\omega_0=2g_1+46g_2$, and the internal of peaks $\delta _{\omega} \sim 8g_2$ for the pseudospin chain with $11$ sites.
    }
\end{figure}

We first consider the situation $g_2=0$, i.e. the McP experiences solely a linear gradient potential.
It is expected that the McP would perform Bloch oscillation in the presence of the linear gradient,
just like a free lattice particle, which provides dynamical evidence that the monopole core and the dressing magnon cloud are indeed bound together and can be treated as one entity, i.e. the composite quasiparticle of McP.
Figures 4(a1) and (b1) present the simulation results with initially a south monopole placed in the middle site,
and the calculation is performed with the original DSG Hamiltonian $\hat H_{\rm{FH}}$. 
In Fig. 4(a1) the temporal density evolution of the one-body density of each site 
$\rho_x(t)=\left\langle \Psi(t) \right| \hat x \left| \Psi(t) \right\rangle$ is shown, which 
clearly shows a periodic evolution and provides the direct signature of the Bloch oscillation.
Moreover, we determine the spectrum of $\rho_x(t)$, as shown in Fig. 4(b1). In the spectrum, a single
intrinsic frequency appears, and it locates exactly at the position of \textbf{$\omega_B=2g_1$},
which coincides with the prediction of the Bloch oscillation.
Figures 4(a1) and (b1) together indicate that the monopole and the dressing magnon cloud indeed move
together as a whole and perform Bloch oscillation under the linear magnetic gradient,
which provides a dynamical confirmation of the validity of the McP picture.

We further investigate the dynamical process of the McP under the quadratic gradient potential
$\hat V^{\rm{McP}}$ with $g_2\neq 0$. It has been revealed that, the lattice particle
under a quadratic gradient undergoes the generalized Bloch oscillation, of which the spectral
fingerprint is the equidistant splitting of the intrinsic frequency \cite{Zhu2019Generalized}. 
The temporal evolution of $\rho_x(t)$ exerted to the quadratic tilt, as shown in Fig. 4(a2),
indicates that the system still undergoes a periodic oscillation,
while the symmetry between the left and right oscillation wings breaks.
The reflection symmetry breaking of $\rho_x(t)$ induces the oscillation of the 
center of mass of the whole system, as illustrated by the red line in Fig. 4(a2).
The corresponding spectrum of $\rho_x(t)$ shown in Fig. 4(b2) presents a series of equidistant
frequency peaks, and the frequency spacing between neighboring peaks obtained numerically with
the original DSG Hamiltonian $\hat H_{\rm{FH}}$ matches $\delta _{\omega} \sim 8g_2$ 
as predicted by the generalized Bloch oscillation of the McP. 
The (generalized) Bloch oscillation of the McP under the linear (quadratic) gradient also illustrates
the controllability of the McP by a magnetic field.

\section{The heterogeneous bipolaron}
 \label{section:IV}
We turn now to an investigation of the coupling effects between different McPs.
Particularly, we focus on the case that one north and one south monopole are simultaneously
doped to the atomic pseudospin chain and form two McPs. The McP with a single monople core of
the north and south type termed as the NMcP and SMcP, respectively.
The basis state of the two-polaron system can be spanned by 
$\{|R,r\rangle=|(x_N+x_S)/2,x_N-x_S\rangle\}$, where $x_{N(S)}$ denotes the position of the
north (south) monopole and $R$ and $r$ correspondingly specify the center-of-mass and
relative coordinates, respectively.
The basis state $|R,r\rangle$ with $r>0$ ($r<0$) denotes the north monopole lying to the 
left (right) of the south one, and $|R,r=0\rangle$ should be eliminated from the basis,
since these states refer to the two monopoles occupying the same site, 
which will lead to the pair annihilation of the two monopoles. 
The pair annihilation, however, provides various second-order channels for exchanging
positions of the two monopoles, i.e. scattering between $|R,r=\pm 1\rangle$, as sketched in Fig. 5(a),
where the intermediate states of these coupling channels are energetically
detuned from the two-monopole basis states.

\begin{figure}[htp]
	\centering
	\includegraphics[trim=20 20 -5 -5,width=0.47  \textwidth]{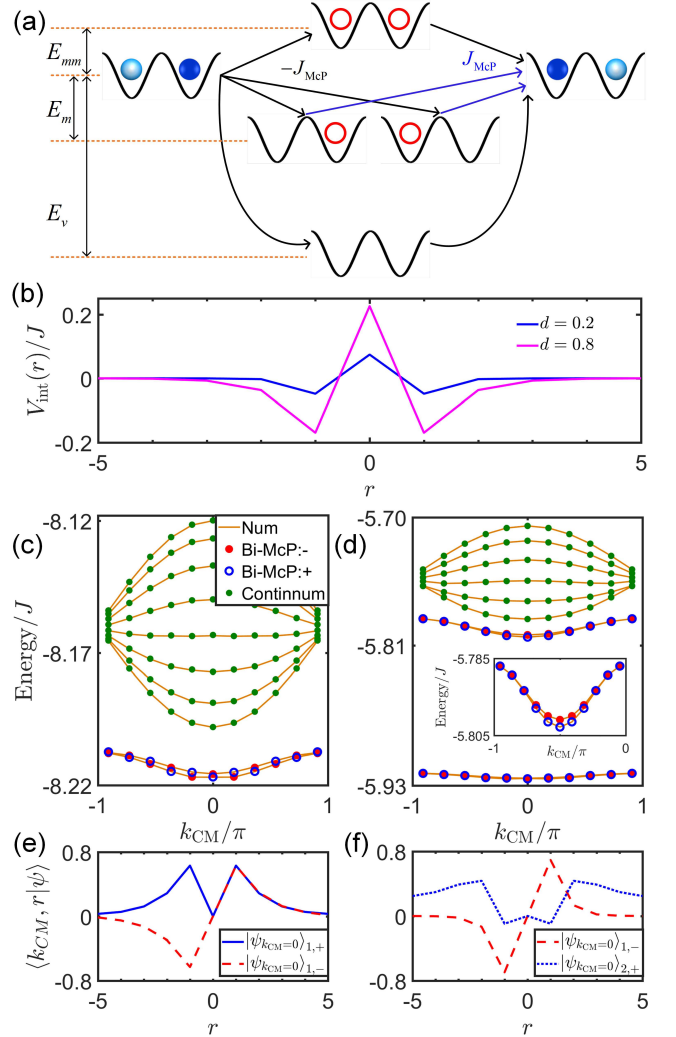}
	\caption{
	\label{fig5}
	(a) Pathway of the second-order coupling for the position exchange of SMcP and NMcP.
	(b) The complete SMcP-NMcP interaction potential $V_{\rm{int}}(r)$ with $d=0.2$ (blue) and $0.8$ (purple).
	(c) and (d) show the spectrum for $d=0.2$ and $0.8$ as a function of the momentum of the center of mass $k_{\rm{CM}}$. 
    The solid circles are determined from the effective model:
    The brown lines indicate the eigenenergy of $\hat H_{\rm{eff}}$.
    The blue hollow circle (red solid circle) and dark green solid circles label the eigenstates with positive (negative) parity in the bipolaron and the quasi-continuum branch.
    The inset in (d) zooms in the second bipolaron band in the main figure.
	(e) and (f) The eigenstates with $k_{\rm{CM}}=0$ with $0.8$ as $k_{\rm{CM}}=0$.
    In (e), the black solid line (red dashed line) refers to the eigenstate with positive (negative) parity of the bipolaron band.
    In (f), the red dashed line (blued dotted line) shows the eigenstate in the first (second) bipolaron band with negative (positive) parity. 
    }
\end{figure}

The two McPs exert nonlocal interaction with each other mediated by the virtual
magnon cloud, and the effective interaction potential $V_{\rm{int}}(r\neq 0)$ can be obtained through the 
Born-Oppenheimer approximation \cite{Dehkharghani2018Coalescence,Will2021Polaron} and taken as 
the ground state energy of the system with the two monopoles fixed at a relative distance of $r\neq 0$,
which can be derived through the Jordan-Wigner transformation.
Despite that $|R,r=0\rangle$ is eliminated from the basis set,
a pseudopotential barrier at $r=0$, i.e. $V_{\rm{int}}(r=0)$, is particularly added to take into
account the coupling between $|R,r=-1\rangle$ and $|R,r=1\rangle$ through the pair-annihilation
channels shown in Fig. 5(a).
The barrier $V_{int}(r=0)$ should induce the same scattering strength as
the pair annihilation channels, which are both second-order processes, and
$V_{int}(r=0)$ is then evaluated as:
\begin{equation}
    \begin{aligned}
    V_{int}(r=0)=(\frac{1}{E_v}-\frac{2}{E_m}+\frac{1}{E_{mm}})^{-1},
    \end{aligned}
\end{equation}
in which $E_v$, $E_m$, and $E_{mm}$ are the energy difference between $|R,r=\pm1\rangle$
and the intermediate states of the vacuum, single-magnon, and two-magnon excitations
in different coupling channels,
respectively. The complete $V_{\rm{int}}(r)$ is plotted in Fig. 5(b), and it can be seen that
the two McPs are subject to an attractive interaction at $r\neq0$, with a 
central repulsive barrier at $r=0$. Figure 5(b) also illustrates that the interaction strength between
the two McPs becomes stronger as $d$ increases, which also confirms the interaction is
mediated by the dressing magnon cloud, since the magnon cloud is more strongly excited
as $d$ increases.

The interaction potential $V_{\rm{int}}(r)$ suggests that the NMcP and SMcP can form a bound state,
which leads to the generation of the emergent bipolaron quasiparticle.
We verify the generation of the bipolaron with the eigenenergy spectrum of the whole system,
as shown in Fig. 5(c), of which the interaction strength is taken as $d=0.2$, corresponding to 
the shallow interaction potential (the blue line) in Fig. 5(b). 
Figure 5(c) shows the spectrum as a function of the momentum of
the center of mass coordinate, which is a good quantum number due to the translation invariance
of the system. The spectrum presents a quasi-continuum and a discrete branch lying below.
The discrete branch is an indicator of the bipolaron, and its appearance in
Fig. 5(c) confirms that the NMcP and SMcP can form a bipolaron with
two distinguishable impurity cores by the attractive interaction.

It is also noticeable in Fig. 5(c) that, the bipolaron branch is two-fold quasi-degenerate,
which is different from the nondegenerate branch of the bipolaron with identical impurity cores.
The two-fold quasi-degeneracy is attributed to the distinguishability of the north and 
south monopole cores in the bipolaron, which effectively bring in an inner degree of freedom (dof) to
the heterogeneous bipolaron. The inner dof is associated to the relative distance $r$, 
and the inner states correspond to the gerade and ungerade superpositions of the bound states
at the local minima of $V_{\rm{int}}(r)$.
Figure 5(d) presents the relative-coordinate wavefunctions of the heterogeneous bipolaron, 
and it can be observed that the wavefunctions indeed present reflection symmetry, with opposite parities.

It can be further expected that more two-fold quasi-degenerate bipolaron branches will appear
in the spectrum as the attractive interaction increases and sustains more bound states in
the local minimum. Indeed, Fig. 5(e) shows that as the interaction strength increases the
eigenenergy spectrum presents two bipolaron branches, each of which is two-fold quasi-degenerate.
In Fig. 5(f), we present the relative-coordinate wavefunctions of the bipolarons in both branches,
where the wavefunctions of the positive and negative parity are for the lower and upper branches,
respectively. It can be found that the wavefunction of the upper branch presents one more node
than that in the lower branch in each local minimum of $V_{\rm{int}}(r)$, confirming the
bipolaron branches in Fig. 5(e) are induced by the binding of the two monopoles by $V_{\rm{int}}(r)$.

\section{Conclusion and outlook}
\label{section:V}

The concept of a polaron originating from condensed matter physics has been generalized to various fields,
such as chemistry \cite{Bredas1985Polarons}, biophysics\cite{Esther2005Charge} and ultracold atomic physics, 
and can further find close connections to various phenomena in particle physics \cite{Kutschera1993Proton,
BISHOP1973391}.
A common property of the polaron in different platforms is the self-trapping effect, which states that
the dressing cloud would suppress the mobility of the impurity core of the polaron. In this work,
we propose a new type of polaron, McP, with the monopole as the impurity core, and reveal the anti-trapping effect
of McP, which is in contrast to the well-known self-trapping effect. The anti-trapping effect of the McP
sheds a new light on the understanding of how the coupling between the impurity core and
the dressing cloud can modify the polaronic properties, which could inspire the control of the polaron by engineering the coupling between the impurity core and the dressing cloud. It is also interesting to ask whether
the anti-trapping effect within the ultracold atomic simulation platform can give a hint to the 
monopole behavior in condensed matter or even in particle physics.

We also propose the heterogeneous bipolaron formed by a pair of indistinguishable north and south monopoles,
and the heterogeneous bipolaron possesses an extra inner degree of freedom comparing to the bipolaron formed
by identical impurity cores. Given that the bipolaron plays an important role in various transport processes,
the extra inner degree of freedom could find potential applications in the manipulation
of the transport properties related to the bipolaron. 

The atomic pseudospin chain with monopole doping can be experimentally realized
within current techniques. The two major ingredients of our proposal are the
double-well superlattice and the NN interaction. The double-well superlattice
can be generated by the superposition of two optical lattices, with the ratio of the wavelength between the long and short lattices
being $\lambda_l/\lambda_s=2$ \cite{Sebby2006Lattice, Anderlini2007Controlled, Folling2007Direct, Martin2016Spin, Salomon2019Direct}.
The NN interaction can be induced by using polar lattice gases \cite{Baranov2012Condensed, Paz2013Nonequilibrium, Yan2013Observation, Baier2016Extended, Li2021Hilbert}, as well as
through the Rydberg dressing. For instance, one can take
$^6\rm{Li}$ atoms as the working medium, for which the double-well superlattice has been
realized by superimposing two standing waves formed by counterpropagating lasers of
wavelength $\lambda _s= 2.3 \ \rm{\mu m}$ and $\lambda _l = 4.6 \ \rm{\mu m}$ 
\cite{Martin2016Spin, Salomon2019Direct}.
The NN interaction can then be realized through the dressing by the Rydberg state
$\left| 34S_{1/2} \right\rangle$ \cite{Gao2022Interaction}.
Quantitatively speaking, taking the amplitudes of short- and long-wavelength lattices
as $20E_R$ and $11.1 E_R$, respectively, will lead to the hopping strength of 
$J=50J_1=97\rm{Hz}$, with the recoil energy $E_R=h/2\lambda _s m_{\rm{Li}}$, 
where $h$ and $m_{\rm{Li}}$ are the Planck constant and the atomic mass.
Provided the van der Waals–type interaction coefficient $C_6=46.5 \ \rm{MHz} \ \rm{\mu m}^6$ 
for $\left| 34S_{1/2} \right\rangle$ of $^6\rm{Li}$, the detuning and Rabi frequency of
the Rydberg excitation laser are chosen as $87.5$ and $5.7$MHz, respectively, to fulfill
$d \sim J$, which can reside the atomic pseudospin chain in the paramagnetic phase.

\section*{Acknowledgements}
The authors would like to acknowledge T.Shi and Y. Chang for inspiring discussions. 
This work was supported by the National Natural Science Foundation of China 
(Grants No. 2022YFA1404102, No. 11625417, No. 11604107, No. 91636219 and No. 11727809),
and the Cluster of Excellence 'Advanced Imaging of Matter' of 
the Deutsche Forschungsgemeinschaft (DFG)-EXC 2056, Project ID No. 390715994.

\newpage

\begin{appendix}

\section{The single monopole-core polaron basis derived  from Jordan-Wigner transformation}
\label{App1}

The derivation of the monopole-core polaron (McP) basis is based on the Born-Oppenheimer approximation.
We consider the motion of monopoles as a slow variable and the density profile of surrounding magnons, $i.e$. the dressing cloud quickly adapts to it.
The McP basis $\left| i \right\rangle _{\Lambda-\rm{McP}}$, which defines an McP induced by $\Lambda$-type monopole at the $i$-th site,
is approximated as (Eq. \ref{eq3} in the main text):
\begin{equation}
    \begin{aligned}
        {\left| i \right\rangle _{\Lambda-\rm{McP}}} = \left| i \right\rangle_{\Lambda} \otimes \left| \psi \right\rangle_{i, \Lambda},
    \end{aligned}
    \label{eqS1}
\end{equation}
where $\left| i \right\rangle _{\Lambda}= \hat \alpha_{i,\Lambda}^\dagger \left| vac \right\rangle$ describes a $\Lambda$-type
monopole located at the $i$-th site, 
$\left| \psi \right\rangle _{i,\Lambda}$ is the ground state of the remaining pseudospin chain.
Here we take the derivation of ${\left| 1 \right\rangle _{S-\rm{McP}}}$ as an example,
and ${\left| i \neq 1 \right\rangle _{S-\rm{McP}}}$ can be obtained by the translational symmetry of DSG system.
The same process can also be applied for the derivation of ${\left| i \right\rangle _{N-\rm{McP}}} $.

\begin{figure}[b]
	\centering
	\includegraphics[trim=30 10 0 0,width=0.46 \textwidth]{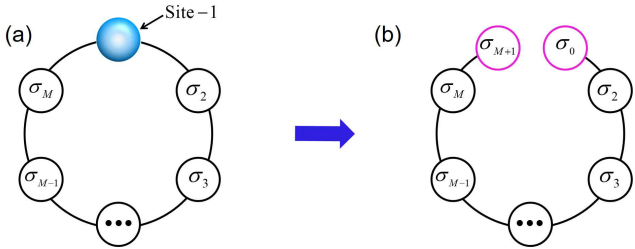}
	\caption{
	\label{figS1}
	(a) The DSG pseudospin chain with an SM fixed at $1$st site.
    (b) The extended pseudo spin chain replacing SM with two auxiliary spins.
    The auxiliary spins are marked with purple circles.
    }
\end{figure}

Figure. \ref{figS1}(a) demonstrates the DSG pseudospin chain with single south magnetic monopole (SM) doping.
The SM is fixed at the $1$st site and pseudospins at the $2$nd and $M$-th sites are affected by its singular magnetic field.
The Hamiltonian of this system is:
\begin{equation}
    \begin{aligned}
        {\hat H_{\rm{S-1}}} =  &2J\sum\limits_{i}^M {\hat m_{i}^\dagger} \hat m_{i}  - \frac{d}{4}\sum\limits_{i}^{M - 1} {\left( \hat m_i ^\dag + {{\hat m}_{i}}  \right)} {\left( {\hat m_{i+1} ^\dag + \hat m_{i+1}} \right)} \\
         & +  \frac{d}{4} {\hat n_{1,S}} \left[\left( {\hat m_{2}^\dag  + {{\hat m}_{2}}} \right) - {\left( {\hat m_{M}^\dag  + {{\hat m}_{M}}} \right) } \right],
    \end{aligned}
    \label{eqS2}
\end{equation}
where $\hat m_{i}^\dagger (\hat m_i)$ is the creation (annihilation) operator of magnon,
${\hat n_{1,S}} \equiv 1$ is the density operator of SM at site-1.
$J$ and $\frac{d}{4}$ refer to the amplitude of the effective transverse magnetic field and monopole-spin (spin-spin) interaction.
The first line in ${\hat H_{\rm{S-1}}}$ describes an Ising spin chain with transverse field, 
and the second line comes from the singular fields of the SM.
On top of that, we also ignore the pair excitation/annihilation of monopoles in Eq.\ref{eq2} of the main text.
${\left| 1 \right\rangle _{S-\rm{McP}}}= \left| 1 \right\rangle_{S} \otimes \left| \psi \right\rangle_{1, S}$
is then the  ground state of ${\hat H_{\rm{S-1}}}$ .

We solve $ \left| \psi \right\rangle_{1, S}$ by introducing two auxiliary spins located at the $0$-th and $M+1$-th site, as shown in Fig. \ref{figS1}(b).
The SM-magnon interaction in Eq.\ref{eqS2} is replaced by the ferromagnetic interaction between auxiliary spins and magnons.
Following the works in Refs \cite{BARIEV1991166, Bilstein_1999, Campostrini2015Quantum, Hu2021First, Hu2021Quantum}, we also require that the auxiliary spins are insensitive to the transverse magnetic field.
The Hamiltonian of the extended pseudospin chain is:
\begin{equation}
    \begin{aligned}
        {\hat H_{\rm{S-1}}}^{\rm{ext}} =  &2J\sum\limits_{i}^M {\hat m_{i}^\dagger} \hat m_{i}  - \frac{d}{4}\sum\limits_{i}^{M - 1} {\left( \hat m_i ^\dag + {{\hat m}_{i}}  \right)} {\left( {\hat m_{i+1} ^\dag + \hat m_{i+1}} \right)} \\
         & + \frac{d}{4} {\left( \hat m_0 ^\dag + {{\hat m}_{0}}  \right)} {\left( {\hat m_{2} ^\dag + \hat m_{2}} \right)} \\
         & - \frac{d}{4} {\left( \hat m_M ^\dag + {{\hat m}_{M}}  \right)} {\left( {\hat m_{M+1} ^\dag + \hat m_{M+1}} \right)}.
    \end{aligned}
    \label{eqS3}
\end{equation}

The Hilbert space of the extended spin chain can be divided into four disconnected sectors 
distinguished by $\left( \left\langle \hat \sigma_{0,z} \right\rangle, \left\langle \hat \sigma_{M+1,z} \right\rangle \right) =  \left( \left\langle \hat m_0^\dagger + \hat m_0 \right\rangle, \left\langle \hat m_{M+1}^\dagger + \hat m_{M+1} \right\rangle \right)$.
The restriction of ${\hat H_{\rm{S-1}}}^{\rm{ext}}$ to sector $\left( 1,-1 \right)$ gives rise to Eq.\ref{eqS2}
and the ground state in this sector is represented as:
\begin{equation}
    \begin{aligned}
        \left| \Psi \right\rangle _{\uparrow \downarrow}=\left| \uparrow \right\rangle _{0} \otimes  \left| \psi \right\rangle_{1, S} \otimes \left| \downarrow \right\rangle _{M+1},
    \end{aligned}
    \label{eqS}
\end{equation}
and therefore we have $\left| \psi \right\rangle_{1, S}= \mathcal{P}_{\uparrow\downarrow}  \left| \Psi \right\rangle _{\uparrow \downarrow} $,
where $\mathcal{P}_{\uparrow\downarrow}$ is the projection onto the subspace $\left( 1,-1  \right)$.

Next we determine the eigenvectors of the quadratic Hamiltonian ${\hat H_{\rm{S-1}}}^{\rm{ext}}$.
We follow the standard Jordan-Wigner transformation \cite{Lieb1961Two} to transform magnons into spinless fermions:
\begin{subequations}
    \begin{align}
        \hat c_j \equiv \exp{\left[ \pi i \sum_{l \neq 1}^{j-1} \hat m_l^\dagger \hat m_j \right]} \hat m_j ,   \\
        \hat c_j^\dagger \equiv \hat m_j^\dagger \exp{\left[ - \pi i \sum_{l \neq 1}^{j-1} \hat m_l^\dagger \hat m_j \right]} ,
    \end{align}
    \label{eqS4}
\end{subequations}
where $\hat c_j ^\dagger$( $\hat c_j$) is the fermion creation (annihilation) operator at site-$j$.
In the spinless fermions representation, ${\hat H_{\rm{S-1}}}^{\rm{ext}}$ takes the form of:
\begin{equation}
    {\hat H_{\rm{S-1}}}^{\rm{ext}}= \sum_{\alpha,\beta=1}^{M+1}\left( \hat c_{\alpha}^{\dagger} {\bf{A}}_{\alpha,\beta} \hat c_{\beta} 
    + \frac{1}{2}\hat c_{\alpha}^{\dagger} {\bf{B}}_{\alpha,\beta} \hat c_{\beta}^{\dagger}
    + \frac{1}{2}\hat c_{\alpha} {\bf{B}}_ {\alpha,\beta} \hat c_{\beta}
    \right),
    \label{eqS5}
\end{equation}
with $\bf{{A}}$ and $\bf{{B}}$ being symmetric and antisymmetric $ (M+1) \times (M+1) $ matrices, respectively.
${\hat H_{\rm{S-1}}}^{\rm{ext}}$ can further be converted into the form:
\begin{equation}
    {\hat H_{\rm{S-1}}}^{\rm{ext}}=  - \frac{1}{2}\sum\limits_{k = 0}^{M} {{\varepsilon _k}}  + \sum\limits_{k = 0}^{M} {{\varepsilon _k}\hat \eta _k^\dag {{\hat \eta }_k}},
    \label{eqS6}
\end{equation}
with the application of the Bogoliubov transformation \cite{Lieb1961Two}:
\begin{equation}
    {\hat \eta _k} = \sum\limits_{\substack{i=0 \\ i \neq 1}}^{M + 1} {\frac{\phi_{k,i}+\varphi_{k,i}}{2}}\hat c_i^\dag  + {\frac{\phi_{k,i}-\varphi_{k,i}}{2}}{{\hat c}_i},
    \label{eqS7}
\end{equation}
where vectors $\phi_k=\left( \phi_{k,0},\phi_{k,2},\phi_{k,3}, \cdots, \phi_{k,M+1} \right)^T$
and $\varphi_k=\left( \varphi_{k,0},\varphi_{k,2},\varphi_{k,3}, \cdots, \varphi_{k,M+1} \right)^T$ satisfy the following equations:
\begin{subequations}
    \begin{align}
         \left( \bf{A}-\bf{B} \right) \phi_k =\Lambda_k \varphi_k,    \\
         \left( \bf{A}+\bf{B} \right) \varphi_k =\Lambda_k \phi_k,
    \end{align}
    \label{eqS8}
\end{subequations}
or
\begin{subequations}
    \begin{align}
        {\bf{C}}  \phi_k= \Lambda_k^2 \phi_k,    \\
        {\bf{D}}  \varphi_k= \Lambda_k^2 \varphi_k.
    \end{align}
    \label{eqS9}
\end{subequations}
where ${\bf{C}}=\left( \bf{A}-\bf{B} \right) \left( \bf{A}+\bf{B} \right)$ and ${\bf{D}=\left( \bf{A}+\bf{B} \right) \left( \bf{A}-\bf{B} \right)}$.
$\bf{C}$ and $\bf{D}$ share the common eigenvalues $\Lambda_k \ge 0$, and $ \varepsilon_k = \sqrt{\Lambda_k}$.
In the general case where $\Lambda_k \neq 0$, $\phi_k$ or $\varphi_k$ is solved from Eq.\ref{eqS9} and
then another one can be obtained from Eq.\ref{eqS8}.
This way the relative sign between $\phi_k$ and $\varphi_k$ is also determined.
For $\Lambda_k = 0$, both $\psi_k$ and $\phi_k$ are determined by Eq.\ref{eqS9}, respectively.
The choice of their relative sign does not affect the energy
but affects the definition of being occupied and unoccupied for this zero-energy mode \cite{Campostrini2015Quantum}.

In the following, we solve the eigenvectors and eigenvalues of the matrix $\bf{C}$. 
For the convenience of discussion, we show $\bf{C}$'s elements for $M=6$:
\begin{equation}
    {\bf{C}} = 4\left(\begin{array}{c|ccccc|c}    
        {{\frac{d^2}{16}}}&{\frac{Jd}{4}}&0&0&0&0&0     \\ \hline
        {\frac{Jd}{4}}&{{J^2} + {\frac{d^2}{16}}}&{\frac{Jd}{4}}&0&0&0&0\\
        0&{\frac{Jd}{4}}&{{J^2} + {\frac{d^2}{16}}}&{\frac{Jd}{4}}&0&0&0\\
        0&0&{\frac{Jd}{4}}&{{J^2} + {\frac{d^2}{16}}}&{\frac{Jd}{4}}&0&0\\
        0&0&0&{\frac{Jd}{4}}&{{J^2} + {\frac{d^2}{16}}}&{\frac{Jd}{4}}&0\\
        0&0&0&0&{\frac{Jd}{4}}&{{J^2} + {\frac{d^2}{16}}}&0\\ \hline
        0&0&0&0&0&0&0
    \end{array}\right).
    \label{eqS10}
\end{equation}

The matrix $\bf{C}$ can be divided into three blocks along its diagonal, as separated by the solid black line:
($i$), ${{C}}(M+1,M+1)$, there is no coupling between $C(M+1,M+1)$ and the other row(column);
($ii$), ${C}(1,1)$, which couples to the adjacent row(column) with $C(1,1) \neq C(2,2)$;
($iii$), the matrix elements not in the first and last rows (column), which is identical to the Hamiltonian of a particle in the box potential.

The trivial eigenvector with zero eigenvalue contributed by block-$i$ is
\begin{subequations}
    \begin{align}
        \phi_0=\left(0,0,\cdots,0,1 \right)^T,    \\
        \varphi_0=\left(1,0,\cdots,0,0 \right)^T,
    \end{align}
    \label{eqS11}
\end{subequations}

There is also another eigenvector representing an edge mode with zero eigenvalue as $M \to \infty$, which is induced by the difference between the diagonal terms of block-$ii$ and $iii$ and their coupling:
\begin{center}
\begin{subequations}
    \begin{align}
        \phi_{1,0}=\frac{1}{\mathcal{N}}  \quad \phi_{1,i}=-\frac{(-1)^i}{\mathcal{N}} (\frac{d}{4J})^i,    \\
        \varphi_{1,L+1}=\frac{1}{\mathcal{N}}  \quad \varphi_{1,L+1-i}=-\frac{(-1)^i}{\mathcal{N}} (\frac{d}{4J})^i,
    \end{align}
    \label{eqS12}
\end{subequations}
\end{center}
\noindent
where $\mathcal{N}$ is the normalized coefficient.
These two zero energy modes originate from different regimes.
The eigenvector described in Eq.\ref{eqS11} is due to the $Z_2$ global symmetry of ${\hat H_{\rm{S-1}}}^{\rm{ext}}$, 
which leads to the degeneracy of the energy spectrum of ${\hat H_{\rm{S-1}}}^{\rm{ext}}$, but not related to the spectrum of ${\hat H_{\rm{S-1}}}$ \cite{Hu2021First}.
The edge mode eigenvector in Eq.\ref{eqS12} can be considered as the defect mode of a single particle in a box potential with an extra pinning trap.

The rest of the $M-1$ eigenvectors are extended states whose eigenvectors and eigenvalue are given by
\begin{subequations}
    \begin{align}
    \phi_{n,0}    &= \frac{1}{\mathcal{N}}{\frac{1}{{{h_L}}}\sin {\theta _n}}, \\
    \phi_{n,i} &= \frac{1}{\mathcal{N}}{\sin \left( {j{k_n} + {\theta _n}} \right)} ,
    \end{align}
\end{subequations}
and
\begin{equation}
    \Lambda_n^2=1+(\frac{d}{4})^2+\frac{d}{4} \cos\left( k_n \right).
\end{equation}
\noindent
where ${\mathcal{N}}$ is the normalized coefficient and 
$k_n=\pi  - \frac{{ {n}\pi  + {\theta _n}}}{{L+1}}$ is the wave vector, 
$\theta _n$ is the phase shift and can be determined by the constraint in the left boundary
\begin{equation}
    J^2{\phi _{n,0}} + \frac{Jd}{4}{\theta _{n,1}} = \varepsilon _n^2{\phi _{n,0}},
\end{equation}
which leads to
\begin{equation}
    \begin{aligned} 
        \cot \left( {{\theta _n}} \right) &= \left( {1 - \frac{32}{{d^2}}} \right)\cot \left( {\frac{{\left(n-1\right)\pi  - {\theta _n}}}{{M + 1}}} \right) \\
        &+ \frac{{d^2/4 - 4 - 4{J^2}}}{{dJ^2}}\frac{1}{{\sin \left( {\frac{{\left(n-1\right)\pi  - {\theta _n}}}{{M + 1}}} \right)}}.
    \end{aligned}
\end{equation}
$\varphi_k$ can be obtained with \ref{eqS8} as
\begin{equation}
{\varphi_k} = \frac{1}{{{\Lambda _k}}}\left( {{\bf{A}} - {\bf{B}}} \right){\phi_k}.
\end{equation}

$\left| \Psi \right\rangle _{\uparrow \downarrow}$ is one of the first degenerate excited state of ${\hat H_{\rm{S-1}}}^{\rm{ext}}$ \cite{Campostrini2015Quantum,Hu2021First},
which reads:
\begin{equation}
    \left| \Psi \right\rangle _{\rm{S-1}}^{\rm{ext}} = \frac{\hat \eta_1^{\dagger}\left( 1+ \hat \eta_0^\dagger \right)}{\sqrt{2}} \prod_{k=0}^M \hat \eta_k \hat \alpha^{\dagger}_{\Lambda,i} \left| vac \right\rangle,
\end{equation}
and the McP basis is then:
\begin{equation}
    \left| \psi \right\rangle _{1,\rm{S}}= \mathcal{P}_{\uparrow,\downarrow} \left| \Psi \right\rangle _{\rm{S-1}}^{\rm{ext}}.
\end{equation}

The mangon density $\rho_{i}^M$ is then:
\begin{equation}
    \begin{aligned}
        \hat n_{i}^M &=\left\langle \Psi \left| \hat c_i^\dagger \hat c_i \right| \Psi \right\rangle \\
        &= \frac{1}{2} (1- \phi_{1,i}\varphi_{1,i} + \sum\limits_{k = 2}^{M} {{\phi_{k,i}}{\varphi_{k,i}}} ).
    \end{aligned}
\end{equation}

\section{The double monopole-core polaron basis derived from Jordan-Wigner transformation}
\label{App2}
\begin{figure}[t]
	\centering
	\includegraphics[trim=25 10 -5 -15,width=0.47 \textwidth]{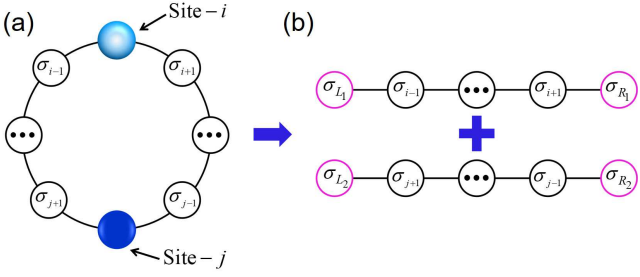}
	\caption{
	\label{figS2}
	(a) The DSG pseudospin chain with one SM (NM) fixed at $i(j)$-th site.
	(b) The two extended sub-pseudospin chains named chain-SN (upper) and chain-NS (lower).
    The auxiliary spins are marked with purple circles.
    }
\end{figure}

This section is a straightforward extension of the Appendix. \ref{App1}.
We apply the same process to obtain the double McP basis where both SM and NM are doped into the pseudospin chain.
As shown in Fig. \ref{figS2}(a), the pseudospin chain is divided into two segments by the SM and NM at $i$-th and $j$-th sites.
The SMcP-NMcP basis is defined as:
\begin{equation}
    \begin{aligned}
        \left| i, j\right\rangle ^{SN}_{\rm{McP}}= \left| i \right\rangle _S \otimes \left| \psi \right\rangle _{i,j}^{SN} \otimes \left| j \right\rangle _N \otimes \left| \psi \right\rangle _{j,i}^{NS}, 
    \end{aligned}
\end{equation}
where $\left| i \right\rangle _{S}= \hat \alpha_{i,S}^\dagger \left| vac \right\rangle$ ( $\left| j \right\rangle _{N}= \hat \alpha_{j,N}^\dagger \left| vac \right\rangle$ )
describes a SM (NM) locates at $i(j)$-th site. 
$\left| \psi \right\rangle _{i,j}^{SN}$ and $\left| \psi \right\rangle _{j,i}^{NS}$ are 
the ground states of pseudospin chain from $i+1$-th to $j-1$-th site and $j+1$ to $i-1$-th site, respectively.

We append two auxiliary spins to the left and right sides of each sub-pseudospin chain to replace the singular magnetic fields of monopoles.
One can also obtain two extended sub-pseudospin chains named chain-SN and chain-NS, as shown in Fig. \ref{figS2}(b),
with the Hamiltonian $\hat H_{S-i,N-j}^{\rm{ext}}$ and $\hat H_{N-j,S-i}^{\rm{ext}}$.
Then $\left| \psi \right\rangle _{i,j}^{SN}$  can be obtained from the ground state of $\hat H_{i-S,j-N}^{\rm{ext}}$ in block $(1,1)_{SN}$,
and $\left| \psi \right\rangle _{j,i}^{NS}$  can be obtained from the ground state of $\hat H_{j-N,i-S}^{\rm{ext}}$ in block $(-1,-1)_{NS}$,
where $(s_{L_1},s_{R_1})_{NS} \left( (s_{L_2},s_{R_2})_{SN} \right)$ is the eigenvalues of $\sigma_{L1} (\sigma_{L2} )$ and $\sigma_{R1} (\sigma_{R2} )$.

\section{The monopole-core polaron basis derived from second-order perturbation treatement}
\label{App3}

In this Appendix, we derive the magnetic polaron basis within a second-order perturbation theoretical treatment.
This method is also based on the Born-Oppenheimer approximation and calculates the ground state of 
the fixed monopole Hamiltonian using perturbation theory.

Here we take the SMcP basis as an example.
The Hamiltonian of the DSG pseudospin chain with the monopole fixed at $i$-th site is:
\begin{equation}
    \begin{aligned}
        {\hat H_{{\rm{S}}-i}} =  &2J\sum\limits_{i}^M {\hat m_{i}^\dagger} \hat m_{i}  - \frac{d}{4}\sum\limits_{i}^{M - 1} {\left( \hat m_i ^\dag + {{\hat m}_{i}}  \right)} {\left( {\hat m_{i+1} ^\dag + \hat m_{i+1}} \right)} \\
         & +  \frac{d}{4} {\hat n_{i,S}} \left[\left( {\hat m_{2}^\dag  + {{\hat m}_{2}}} \right) - {\left( {\hat m_{M}^\dag  + {{\hat m}_{M}}} \right) } \right],
    \end{aligned}
    \label{eqSC0}
\end{equation}
which can be decomposed into ${\hat H_{{\rm{S}}-i}}={\hat H_{{\rm{S}}-i} ^{(0)}}+{\hat H_{{\rm{S}}-i} ^{(1)}}$:
\begin{subequations}
    \begin{align}
        {{\hat H_{{\rm{S}}-i} ^{(0)}}} =  &2J\sum\limits_{i}^M {\hat m_{i}^\dagger} \hat m_{i}  \\
        {{\hat H_{{\rm{S}}-i} ^{(1)}}} =- &\frac{d}{4}\sum\limits_{i}^{M - 1} {\left( \hat m_i ^\dag + {{\hat m}_{i}}  \right)} {\left( {\hat m_{i+1} ^\dag + \hat m_{i+1}} \right)} \\
         +  &\frac{d}{4} {\hat n_{i,S}} \left[\left( {\hat m_{2}^\dag  + {{\hat m}_{2}}} \right) - {\left( {\hat m_{M}^\dag  + {{\hat m}_{M}}} \right) } \right],
    \end{align}
\end{subequations}
where ${\hat H_{{\rm{S}}-i} ^{(0)}}$ is the chemical potential of the magnon which comes from the transverse magnetic field,
${\hat H_{\rm{S-1}} ^{(1)}}$ describes the SM-spin and spin-spin interaction.
In the paramagnetic phase, ${\hat H_{{\rm{S}}-i} ^{(0)}}$ plays the leading rule and ${\hat H_{\rm{S-1}} ^{(1)}}$ can be considered as a perturbation.

The ground state of ${\hat H_{{\rm{S}}-i} ^{(0)}}$, $i.e.$ the zeroth order wavefunction of McP
is the vacuum state of magnons $\left| i \right\rangle _{S-\rm{McP}}^{(0)} = \hat \alpha_{i,S}^\dag \left| {Vac} \right\rangle$,
with the ground eigenenergy $E_{0}^{(0)}=0$.
The excited bands are characterized by the number of magnons 
and there are $n$ magnons in the $n$-th excited band with the corresponding eigenenergy $2nJ$. 

In first-order perturbation, the states $\hat m_{i \pm 1}^\dagger \hat \alpha_{i,S} \left| vac \right\rangle$ are in the first excited band and \\
$\sum _{j \neq i,i-1} \hat m_{j}^\dagger \hat m_{j+1}^\dagger \hat \alpha_{i,S} \left| vac \right\rangle$ in the second excited band which are coupled to $\left| i \right\rangle _{S-\rm{McP}}^{(0)}$,
as a result of the monopole's singular magnetic fields and spin-spin interaction.
The McP basis is obtained as
\begin{equation}
    \left| i \right\rangle _{S-\rm{McP}}^{(1)} = \frac{1}{\mathcal{N}_1 }\left\{ \begin{array}{l}
        1 - \frac{d}{8}\left( {\hat m_{i - 1}^\dag  - \hat m_{i + 1}^\dag } \right) \\
        + \frac{d}{16}\sum\limits_{j \ne i,i-1} {\hat m_j^\dag \hat m_{j + 1}^\dag }    \\
    \end{array}    \right\}\hat \alpha_{x,S}^\dag \left| {Vac} \right\rangle ,
    \label{EqSC1}
\end{equation}
We apply a normalization to $\left| i \right\rangle _{S-\rm{McP}}^{(1)}$ with normalization coefficient $\mathcal{N}_1$.

In the second-order perturbation, the basis which has up to $4$ magnons is coupled to $\left| i \right\rangle _{S-\rm{McP}}^{(0)}$ through the second-order process,
and we have:
\begin{equation}
    \left| i \right\rangle _{S-\rm{McP}}^{(2)} = \frac{1}{\mathcal{N}_2 }\left\{ \begin{array}{l}
        1 - \frac{d}{8}\left( {\hat m_{i - 1}^\dag  - \hat m_{i + 1}^\dag } \right) \\
        - \frac{{3{d^2}}}{{64}}\left( {\hat m_{i - 2}^\dag  - \hat m_{i + 2}^\dag } \right)  
        - \frac{{{d^2}}}{{256}}\hat m_{i - 1}^\dag \hat m_{i + 1}^\dag   \\
        - \frac{{{d^2}}}{{192}}\sum\limits_{j \ne i} {\left( {\hat m_{i - 1}^\dag  - \hat m_{i + 1}^\dag } \right)\hat m_j^\dag \hat m_{j + 1}^\dag } \\
        + \frac{d}{16}\sum\limits_{j \ne i} {\hat m_j^\dag \hat m_{j + 1}^\dag }  + \frac{{{d^2}}}{{512}}\sum\limits_{j \ne i} {\hat m_j^\dag \hat m_{j + 2}^\dag }   \\
        + \frac{{{d^2}}}{{256}}\sum\limits_{j,k \ne i} {\hat m_j^\dag \hat m_{j + 1}^\dag \hat m_k^\dag \hat m_{k + 1}^\dag } 
    \end{array} \right\}\hat \alpha_{x,i}^\dag \left| {Vac} \right\rangle, 
    \label{EqSC2}
\end{equation}
where $\mathcal{N}_2$ are the normalized coefficients.
The basis with odd and even numbers of magnon excitations are related to the monopole-magnon interaction and spin-spin interaction, respectively.

One can also apply the same process to obtain the McP basis with more monopoles.

\end{appendix}

\bibliographystyle{apsrev4-2}
\bibliography{refs}

\end{document}